\documentclass{iau}
\usepackage{graphicx}

\title[Census of Nearby Debris Disk Stars] 
{Comprehensive Census and \\ Complete Characterization of \\ Nearby Debris Disk Stars}

\author[T. H. Cotten \& I. Song]   
{Tara H. Cotten$^1$
 \and Inseok Song$^1$}

\affiliation{$^1$University of Georgia\\email: {\tt tara@physast.uga.edu}\\
email: {\tt song@uga.edu}}

\pubyear{2015}
\volume{314}  
\pagerange{119--126}
\jname{Young Stars \& Planets Near the Sun}
\editors{J. H. Kastner, B. Stelzer, \& S. A. Metchev, eds.}
\begin{document}

\maketitle

\begin{abstract}
Debris disks are intimately linked to planetary system evolution since the rocky material surrounding the host stars is believed to be due to secondary generation from the collisions of planetesimals.
With the conclusion and lack of future large scale infrared excess survey missions, it is time to summarize the history of using excess emission in the infrared as a tracer of debris and exploit all available data as well as provide a comprehensive study of the parameters of these important objects.
We have compiled a catalog of infrared excess stars from peer-reviewed articles and performed an extensive search for new debris disks by cross-correlating the Tycho-2 and AllWISE catalogs.
This study will conclude following the thorough examination of each debris disk star's parameters obtained through high-resolution spectroscopy at various facilities which is currently ongoing.
We will maintain a webpage (www.debrisdisks.org) devoted to these infrared excess sources and provide various resources related to our catalog creation, SED fitting, and data reduction.
\keywords{stars: evolution - circumstellar}
\end{abstract}

\firstsection 
\section{Introduction}\label{sec:intro}
The InfraRed Astronomical Satellite (\emph{IRAS}) laid the foundation for large-scale surveys of infrared excess detections in the 1990s following the discovery of the debris disk around Vega (\cite[Aumann et. al. 1984]{aumann_1984}).  
This type of excess emission in the infrared maps the debris in regions analogous to the Kuiper and Asteroid belts.  
With the completion of the most recent all-sky infrared mission (WISE; \emph{Wide-Field Infrared Survey Explorer}, Wright et al. 2010) and without the advent of new dedicated surveys, the amalgamation of all infrared data today will serve as a collective legacy for the detection of debris disk stars.
In addition, as technology advances to directly detect planets embedded in mature stellar systems, the result of this study will provide ideal targets for exoplanet investigations.
The goal of this study is to compile the most comprehensive catalog of high-fidelity infrared excess stars through a combination of prior studies and new detections using the Tycho-2 catalog and the recent release of the WISE catalog (AllWISE).

Though there have been hundreds of debris disk stars published in the literature including those from WISE data searches (i.e. \cite[Wu et al. 2013]{wu_2013}, \cite[Patel et al. 2014]{patel_2014}), only a small fraction of those objects have comprehensive stellar and disk information necessary for uniting theories of planetary formation to observations.  
To thoroughly explore the relationship between disks and a planetary system, stellar information such as effective temperature, metallicity, rotational velocity, radial velocity, and age indicators (CaII H \& K, Li 6708 \AA{}) using optical spectroscopy must be obtained for the catalog of infrared excess stars provided by Cotten \& Song (2015, in prep.).
In particular, one such relationship is based on the trend found in which higher mass planet host stars have higher metallicities (\cite[Fischer \& Valenti 2005]{fischer_2005}).
\cite{greaves_2006} examined a limited sample of debris disk stars to link the dusty disk to the metal-rich star but were unable to recognize any trend.
Moreover, \cite{trilling_2007} and \cite{rodriguez_2012} investigated whether multiplicity would promote or hinder planetary system formation and evolution, especially since stars prefer to form in multiples (\cite[Patience et al. 2008]{patience_2008} and reference therein).
While many factors play a role in these systems, \cite{trilling_2007} and \cite{rodriguez_2012} produced contradictory results regarding whether debris is common among multiple star systems.
Finally, the connection between the rotation of the host star and the surrounding dusty debris disk has only recently been examined and remains an open question.
Specifically, \cite{sierchio_2010} and \cite{mizusawa_2012} outlined contrasting findings, however, the use of $v\sin{i}$ as an indicator for rotational velocity has an inherently large uncertainty due to the inclination angle.

The literature concludes that the connection between stellar properties and dust evolution can only be fully confirmed with larger samples and accurate stellar information.
To aid in this endeavor, the creation of the complete census of infrared excess stars from the literature and a new Tycho-2 and AllWISE correlation is described in Section \ref{sec:census}.
Section \ref{sec:future} itemizes the ongoing observations and future plans to investigate the relationships described above with the largest sample of debris disk stars to date.

\begin{figure}[b]
\begin{center}
\includegraphics[width=2.4in]{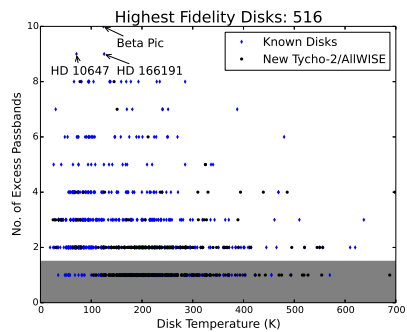}
 \caption{(\emph{Left}) Number of passbands in the infrared which display excess emission.  The stars that have excess confirmed by more than one wavelength make up the high fidelity infrared excess sample.}
   \label{fig:numexcess}
\end{center}
\end{figure}

\section{Creation of Census}\label{sec:census}
The creation of the complete census of debris disk stars began with the selection of a couple of survey publications involving infrared excess (\cite[Rhee et al. 2007]{rhee_2007} and \cite[Rieke et al. 2005]{rieke_2005}) and a review paper by \cite{wyatt_2008}.  
Citations by and of these articles amassed 193 papers with 863 unique sources of infrared excess. 
Given the improvements to infrared excess detections by more recent surveys such as WISE, we re-evaluated the amount of excess emission above the photosphere including all available photometry as well as inspected the AllWISE catalog images to remove sources that may be contaminated by nearby stars, extended cirrus contribution, or background infrared sources.  
In addition, since most of the previously known excess sources are \emph{Hipparcos} stars, we used the accurate parallax measurements to only maintain stars within 125 pc.
The selections from the literature constitutes 530 stars.

Generating the largest census would be incomplete without a full exploitation of the AllWISE catalog containing infrared photometry at 3.5, 4.6, 12, and 22$\mu$m for over 740 million sources.
Therefore, we chose the Tycho-2 catalogue of the 2.5 million brightest stars (\cite[H$\o$g et al. 2000]{hog_2000}) as a compatible match given the reliable optical photometry and proper motions.
Lacking accurate \emph{Hipparcos} parallax measurements, we used the total proper motion ($\mu_{tot}$) as a proxy for distance and selected sources with $\mu_{tot} > 25 $mas/yr corresponding to within 200 pc.
These stars cross-correlated with the AllWISE catalog using a five arc second search radius returned over 513,000 stars.
A number of further criteria were implemented to ensure high quality photometry for SED fitting, that we can eliminate as many giants as possible since only \emph{Hipparcos} stars have precise distance measurements, and finally that the stars show a significant amount of excess above the photosphere.
The summary of these criteria are included in Table \ref{tab:selectioncriteria} and more thorough description can be found in Cotten \& Song (2015, in prep).

\begin{figure}[b]
\begin{center}
\includegraphics[width=2.4in]{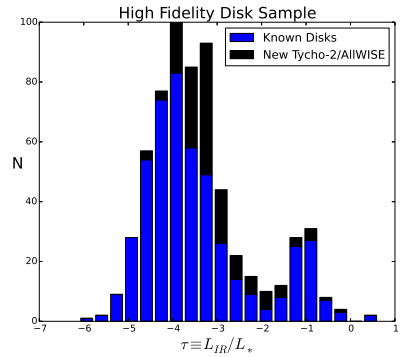} 
 \caption{(\emph{Right}) Histogram of the fractional dust luminosity for our high fidelity sample of infrared excess stars. The secondary peak of significantly dusty sources is comprised of T-Tauri and Herbig Ae/Be objects, although recent investigation found a handful of extremely dusty debris disk stars which are currently under further investigation.}
   \label{fig:tauhist}
\end{center}
\end{figure}
Ruling out any duplicates found between the literature search and this new Tycho-2/AllWISE search, assembled 963 unique infrared excess stars.  
Assessment of the quality of the infrared excess detection required the development of a measure portrayed in Figure \ref{fig:numexcess} defined to be the number of infrared passbands displaying excess emission above the photosphere.
Since we have performed extensive SED fitting using Tycho-2 B and V, \emph{2MASS} J, H, and K$_{S}$, and AllWISE W1 and W2 to define the stellar photosphere, we then examined the amount of excess for all available infrared data which could include AllWISE, IRAS, MSX, Akari, \emph{Spitzer}, and \emph{Herschel}.
As can be seen in Figure \ref{fig:numexcess}, there are a number of stars having excess in only one passband (usually AllWISE W4) in the grey region.
These stars will be maintained in our study, but the stars outside of this region (516 unique sources) display the highest fidelity infrared excess through confirmation by an additional passband. 
Lastly, Figure \ref{fig:tauhist} displays the fractional dust luminosity ($\tau$) for the highest fidelity sample demonstrating there are two populations of dustiness where the extremely dusty disks are mostly comprised of young T-Tauri or Herbig Ae/Be stars, however, a handful of stars found through our new method show significant amounts of dust and appear to be true debris disk stars.

\begin{table}
  \begin{center}
  \caption{Summary of High Fidelity Infrared Excess Star Selection Criteria}
  \label{tab:selectioncriteria}
 {\scriptsize
  \begin{tabular}{|c|l|c||}\hline 
{\bf Selection Criteria} & {\bf No. of Stars }\\ & {\bf Remaining} \\ \hline
AllWISE Catalog & $\sim$ 740 million \\
Total Proper Motion for Dist. Proxy $>$ 25 mas/yr & 515518 \\
5.0 arcsec Cross-Correlation between Tycho-2 and AllWISE & 513478 \\
Well-Measure AllWISE photometry (no upper limits) & 263955 \\
HIP CMD and various color-color diagrams to exclude giants & 245874 \\
Reliable Goodness-of-Fit ($\chi^2$) from SED fitting & 243584 \\
Significance of Excess ($>5\sigma$ in W3 \emph{OR} W4 in each temperature division) & 4537 \\
Image Analysis and Inspection to remove contaminators & 1030 \\
New Infrared Excess Stars within 125 pc & 595 \\ \hline
Known Infrared Excess Stars within 125 pc in Literature & 530 \\ \hline
Final Combined Sample of High Fidelity Infrared Excess Stars (excluding duplicates) & 516\\ \hline
  \end{tabular}
  }
 \end{center}
\vspace{1mm}
\end{table}

\section{Future Work}\label{sec:future}
Characterization of the complete catalog of infrared excess sources will enable a full analysis of the relationship between a star, disk, and potential planets and improve upon past studies by expanding the number of debris disk stars by more than three times.
The parameters we seek to obtain include age, metallicity, surface gravity, stellar rotation, radial velocity, and multiplicity.
Majority of these parameters can only be obtained through high-resolution optical spectroscopy and so we searched the spectral archives of instruments such as FEROS, ELODIE, SOPHIE, HARPS, and UVES so we do not re-observe these stars.
Due to nearly half of the final sample lacking archival spectra, we have distributed sources using numerous parameters such as declination, brightness, and dustiness to collaborators across the globe as well as proposed for additional telescope time at various facilities.
Our census may combine a heterogeneous collection of spectra, but we plan to derive the aforementioned stellar parameters in a uniform manner.
In addition, following the publication of the complete census (Cotten \& Song 2015), we will make our catalog publicly available through the webpage: http://www.debrisdisks.org.
The webpage will serve not only as a catalog of photometric and spectroscopic information, it will also contain interactive plotting capabilities for the reduced spectra as well as our Python SED fitting algorithm.

\begin{discussion}

\discuss{Metchev}{In your plot of significance of excess histograms, how many stars are at -5$\sigma$?}

\discuss{Cotten}{I can't tell you the exact number, but I know it is significantly less than that the number of excess sources.}

\end{discussion}

\end{document}